# Loss Rate Based Fountain Codes for Data Transfer

Jianxin Liao, Lei Zhang, Xiaomin Zhu, Jingyu Wang
State Key Lab of Networking and Switching Technology
Beijing University of Posts and Telecommunications
Beijing, China 100876
Email: {liaojianxin, zhanglei2, zhuxiaomin,
wangjingyu}@ebupt.com

Minyan Liao
Department of Content Research
Creative Knowledge Ltd.
Beijing, China 100089
Email: sophie_hi@qq.com

*Abstract*—Fountain codes are becoming increasingly important for data transferring over dedicated high-speed long-distance network. However, the encoding and decoding complexity of traditional fountain codes such as LT and Raptor codes are still high. In this paper, a new fountain codes named LRF (Loss Rate Based Fountain) codes for data transfer is proposed. In order to improve the performance of encoding and decoding efficiency and decrease the number of redundant encoding symbols, an innovative degree distribution instead of robust soliton degree distribution in LT (Luby Transfer) codes is proposed. In LRF codes, the degree of encoding symbol is decided by loss rate property, and the window size is extended dynamic. Simulations result using LRF codes show that the proposed method has better performance in term of encoding ratio, degree ratio, encoding and decoding efficiency with respect to LT and Raptor codes.

## I. Introduction

With the development of internet, network becomes faster and more reliable. The backbone of china internet constructed by DWDM（Dense Wavelength Division Multiplexing）system has reached to 40Gb/s per single fiber. CMCC (China Mobile Communications Corporation) has established a test system of DWDM from Hangzhou to Fuzhou which is reached to 100Gb/s per single fiber. Some manufacturers have already started to test 400Gb/s per single fiber, even 1Tb/s. As the internet evolves to include many very high speed and long distance network paths, the time spent to transfer large data is an important factor of many systems. For examples, CDN (Content Delivery Network) systems interconnected by high-speed long-distance network need to transfer data from source nodes to cache nodes. Cloud computing systems interconnected by high-speed long-distance network need to replicate data between each other in order to get ability to protect from disaster. However, the performance of traditional TCP (Transmission Control Protocol) data transfer mechanism was challenged in high-speed long-distance network [1]. TCP ensures reliability by essentially retransmitting symbols within a sliding window when three duplicate acknowledge are received or timeout is detected. For instance, if send received three duplicate acknowledge, the send window is decreased to half. If the bandwidth of a network path is 10Gbps and the RTT is 100ms, with symbols of 1250 bytes, TCP takes about 50,000 RTTs (1.4 hours) to full-utilize the path.

Digital fountain codes are spare-graph codes developed for erasure channels, and their key property is that the source data can be recovered from any subset of the encoded symbols, given that enough symbols are received [2]. The first universal fountain codes named LT (Luby Transfer) codes is proposed by Luby in 2002 [3]. Although LT codes demonstrates excellent efficiency, but the decoding cost of it is still too high to be afforded in practical applications. Raptor codes is proposed by Shokrollahi, which is an extension of LT codes with linear time encoding and decoding [4]. In LT and Raptor codes, encoding symbols are generated without knowledge of loss rate, and thus the total degree of encoding symbols should be large enough to cover all input symbols. But when data is transferred over high-speed long-distance with lower loss rate, the raptor codes transfer lots of encoding symbols which is not necessary.

In this paper, we assume that lost symbols are record at destination and loss rate is sent back to source from destination. Since the loss rate is changed in the process of transmission, traditional forward error codes is not suitable. Thus, a new fountain codes named LRF (Loss Rate Based Fountain) codes for data transfer in high-speed long-distance network is introduced. In the proposed LRF codes, an innovative degree distribution instead of robust soliton degree distribution in LT codes is proposed. The key feature of LRF codes is to design a degree distribution to reduce the total number of encoding symbols and total degree of all encoding symbols. In LRF codes, the degree of encoding symbol is decided by loss rate. This approach guarantees that a lost symbol is recovered by an encoding symbol with high probability. It also reduces the total number of encoding symbols, and guarantees that the total degree of all encoding symbols is small too.

Using LRF codes as a transfer codes has two advantages. First, in TCP-based transfer protocol, lost symbols in high-speed long-distance network need to be retransmitted. The send window size is reduced, and then the throughput of transfer is greatly reduced. By using LRF codes, lost symbols can be recovered at receiver by encoding symbols which have been received before. Second, fountain codes such as LT codes and Raptor codes are more complex and the encoding and decoding process need more memory and CPU time. In some situation, the encoding and decoding process is the bottleneck of data transfer system. Data transfer protocol based on LRF codes obtains higher throughput and spends less memory and CPU than LT codes and Raptor codes.

The rest of the paper is organized as follows. In section Ⅱ, background of fountain codes is given. In section Ⅲ, an overview of LRF codes is demonstrated. In section Ⅳ, a mathematical model of degree distribution is presented, and LR-Raptor codes are proposed. Additional, performance of LRF codes is analyzed. In section Ⅴ, our experimental design is outlined, and efficiency of LRF and LR-Raptor codes is illustrated by experiment results. Finally, our work and prospect future work are summarized.

## II. BACKGROUND

The links in network is often modeled as an erasure channel which symbols may be erased with a probability $p$. This channel was introduced by Elias [5] in 1955. The general problem of transfer data based on coding is well studied. Byers et al proposed the idea of digital fountain codes in 1998 [6].

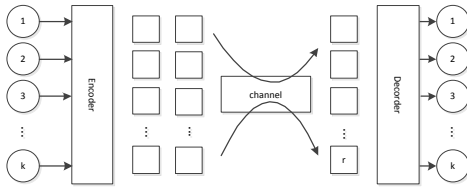

Fig. 1. Fountain codes

As show in Fig. 1, the encoder of fountain codes is like a fountain spewing. Infinite coded symbols can be produced. Source data is divided into $k$ input symbols of size $l$. With fountain codes, the $k$ input symbols are combined into infinite encoding symbols at source. All $k$ input symbols can be recovered from any set of $(1+\varepsilon) k$ encoding symbols, where $0<\varepsilon<<1$. Encoder of fountain codes is bit rate independent which is not limited by the size of the source data and can generate an unlimited number of encoding symbols.

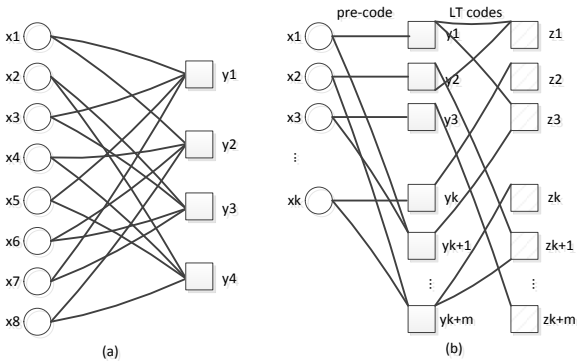

Fig. 2. (a)Tornado codes; (b) Raptor codes

Luby, et al introduced a simple erasure recovery algorithm for codes with linear time encoding and decoding. But they need carefully designing graphs to construct for linear codes [7]. Tornado codes are erasure block codes based on irregular spare graph. Given an erasure channel with loss probability $p$, they can correct up to $p \cdot (1-\varepsilon)$ errors. They can be encoded and decoded in time proportional to $n \cdot \log (1/\varepsilon)$. As shown in Fig. 2(a), there are eight input symbols named $x_1, x_2 \ldots x_8$. With tornado codes, four encoding symbols named $y_1, y_2, y_3$ and $y_4$ is produced by eight input symbols. Tornado codes can tolerate that any one of $y_1, y_2, y_3$ and $y_4$ can be recovered by three others. However, the complexity of encoding and decoding algorithms for tornado codes is proportional to block-length. This makes tornado codes not be adequate for large data transfer systems.

Ruby proposed LT codes which is the first implementation of digital fountain codes in 2002 [3]. With LT codes, data was divided into fix size blocks. Each block is divided into fix size symbols. So the number of input symbols is fixed. Infinite coded symbols can be generated by encoder of LT codes. All input symbols can be recovered by decoder in LT codes when number of encoding symbols are received slightly larger than number of input symbols. Digital Fountain, Inc. proposed Raptor codes in 2006 [4]. It is a concatenation of a systematic pre-code with LT codes. As shown in Fig. 2(b), in the pre-code, $k$ native symbols are first mapped to $(1+\varepsilon) k$ pre-coded symbols. Infinite coded symbols can be generated from pre-coded symbols by LT codes. In decoding process of Raptor codes, pre-coded symbols are recovered by LT codes firstly, and then input symbols are recovered by pre-coded symbols. Raptor code has been standardized in the 3GPP (Third Generation Partnership Project) [8]. A fixed-rate error control fountain codes was proposed by Sivasubramanian, et al [9]. In this fountain codes, a global decoding algorithm, incorporating feedback between the component codes of the Raptor code, is introduced to improve the performance over both memoryless and correlated fading channels. But there is unfair in shared network while new flow is entered into network. The performance of Raptor codes over $F_q$ is investigated in [10]. This paper shows that higher order Galois field is beneficial, in terms of performance. A novel fountain coding scheme with non-binary LDPC codes was proposed by Kasai, et al [11]. Both [10] and [11] reduce the decoding complexity. An UEP (Unequal Error Protection) rateless codes was proposed by Rahnavard, et al [12]. In UEP rateless codes, input block is partitioned into high-priority part and low-priority part. In [13], an UEP method for streaming media was presented. This method guarantees that high-priority input data are recovered before low-priority ones, and important information can be recovered with low latency, low computation, and high probability. A sliding-window digital fountain codes was proposed by Bogino, et al [2], and sliding-window raptor codes was proposed by Cataldi, et al [14]. In order to enhance the performance, they both extend the block length by sliding window. But the complexity of encoding and decoding is still high. A window fountain code was proposed in [15]. In this paper, an alternative way to extend the idea of LT codes is by assigning the set of message symbols into a sequence of (possibly overlapping) subsets. Each subset is usually consecutive and this subset is called as a window. Within each window, the information symbols are usually drawn uniformly to produce an encoding symbol. However, windowed fountain codes incur higher decoding complexity. Expanding Window

Fountain (EWF) codes was introduced by Sejdinovic, et al [16]. In EWF codes, a predefined sequence of strictly increasing subsets of the information symbols are chosen to be a sequence of windows, the input symbols in the smallest window will have the strongest error protection, because they are also contained in all larger windows and are more likely to be used when producing encoding symbols. However, the encoding and decoding complexity is still high.

### III. LRF CODES

In LT codes, encoding symbols are generated according to the coding graph. As shown in Fig. 3(a), the edges between input symbols and encoding symbols are randomly generated using soliton distribution. However, there is only small number of symbols lost while data is transferred over dedicated high-speed long-distance network. In destination, only small number encoding symbols (slightly more than number of lost symbols) can cover all lost symbols. As shown in Fig. 3(b), we assume that one input symbol will be lost by channel, so only one encoding symbol with degree four is generated to recover the lost symbol. The main idea of proposed scheme is to virtually decrease the average degree of codes, so as enhance the performance of the fountain code by reducing the encoding and decoding complexity and increasing transmission efficiency.

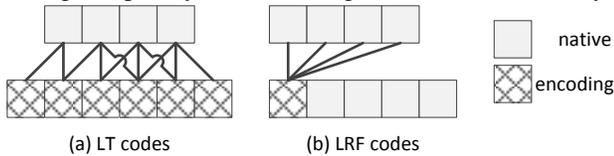

Fig. 3. LT codes versus LRF codes

In LT codes, the block-based partitioning of the data is adopted and represented in Fig. 4(a). Let $k$ be the length of the source block to be encoded, the receive window can slide after received $(1+\varepsilon)k$ encoding symbols. In LRF codes, the window size is dynamic adjusted by transmission layer protocol. The receive window can slide after symbols arrive in perfect order and, more than that, the average degree is less which reduces the complexity of encoding and decoding.

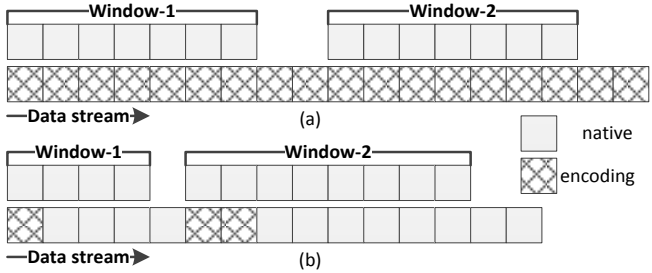

Fig. 4. Fixed Windows versus Dynamic Windows

In order to demonstrate the combination of LRF codes with transmission layer protocol, an application of transferring a large data from source to destination is proposed. As shown in Fig. 5, there is a bi-directional connection. Data is sent from source to destination, and acknowledgment is sent from destination to source. In the source point, feedback is received from destination, and window size and encoding symbols are determined by those feedback information. In order to eliminate the influence of performance reducing caused by native symbols which are lost in transmission channel, encoding symbols are sent in advance to recover the lost symbols. In the destination point, encoding symbols are received firstly and are stored in memory or disk. When the loss of a native symbol is detected, encoding symbols are examined. If this lost native symbol can be recovered by one of encoding symbols, receiver window is slide and an acknowledgment is sent to source, then this encoding symbol is marked as used. Information of feedback is sent to source when loss rate increased or decreased. The detail of combination of LRF codes with transmission layer protocol will be demonstrated in other papers. In this paper we will not address these and other applications, but will instead focus on the theory of LRF codes.

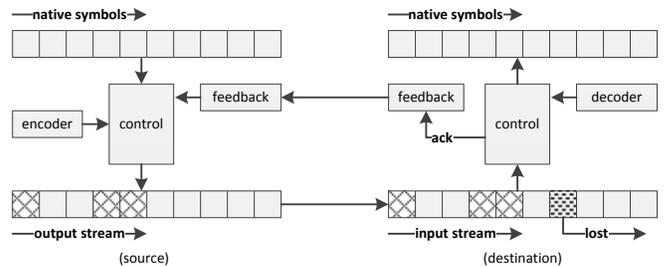

Fig. 5. LRF codes with Transmission Layer Protocol

### IV. DEGREE DISTRIBUTION DESIGN

Suppose we want to transmit a message comprising of $w$ symbols $I = s_1, s_2, \ldots, s_w$ from source to destination. After $w$ input symbols are transmitted, $n$ symbols $O_1 = s'_1, s'_2, \ldots, s'_n$ are obtained at the destination and $m=w-n$ symbols are lost by transmission channel. In LRF codes, instead of automatic repeat transmitting $m$ lost symbols, $m' = (1+\varepsilon)m$ encoding symbols $O_2 = s''_1, s''_2, \ldots, s''_{m'}$ is generated and transmitted to destination in advance. In destination, lost symbols are determined by encoding symbols. In general, the goal of degree distribution design is to minimize the number of $m'$ encoding symbols which are required to give a high probability of decoding $m$ lost symbols.

#### A. Analysis of Minimum Degree

In this subsection, we examine the minimum degree under iterative decoding. In decoding process, $n$ input symbols are initially covered, and the set of covered input symbols that have not yet been processed is called the ripple, and thus at this point all $n$ input symbols are in the ripple. Every input symbol in the ripple is removed as a neighbor from all encoding symbols which have it as a neighbor. Ripple is grown, if one encoding symbol that have exactly one remaining neighbor is released to cover input symbols previously uncovered. If ripple is empty after all $n$ input symbols have been processed, the encoding process is failed. Let $d$ denotes the degree of an encoding symbol. It is clear from the previous process that the success determined an uncovered symbol requires an encoding symbol which has an uncovered symbol as a neighbor and *d*-1 covered symbols as neighbors. The analysis of the classic balls and bins process can be applied in this situation. The number of combinations of selecting *d* symbols from *w* symbols is $C_w^d$. The number of combinations of selecting a symbol from *m* uncovered symbols and selecting *d*-1 symbols from *n* covered symbols is $C_m^1 C_n^{d-1}$. The probability of an uncovered symbol is success determined from an encoding symbol with degree *d* is shown in formula (1). When $d = \frac{w}{m}$, the probability $\rho_{(1)}$ reaches maximum.

$$\rho_{(1)} = \frac{C_m^1 C_n^{d-1}}{C_w^d} \quad d = 1,2,\ldots, n+1. \quad (1)$$

$$d = \frac{w}{m}. \quad (2)$$

Proof:

$$\rho_{(i)} = \frac{C_m^i C_n^{d-i}}{C_w^d} \quad d = 1,2,\ldots, n+1. \quad (3)$$

$$1 = E_{(i)} = \frac{dm}{w}. \quad (4)$$

Let $\rho_{(i)}$ shown in formula (3) denotes probability of *i* uncovered symbols is selected when select *d* symbols from *w* symbols, and $\rho_{(i)}$ obeys hypergeometric distribution. The mathematical expectation of $\rho_{(i)}$ is shown in formula (4). By set $E_{(i)} = 1$, we get $d = \frac{w}{m}$.

*B. Ideal Degree Distribution*

In traditional fountain codes, such as LT codes, the block with *w* input symbols can be decoded if at least (1+ε) *w* encoding symbols are received. In LRF codes, *w* input symbols can be decoded if output symbols contained *n* native symbols and at least (1+ε) *m* encoding symbols. In LT codes, the basic property required of a good degree distribution is that encoding symbols are added to the decoder as the same rate as they recovered out input symbols. In LRF codes, a desired effect is that an uncovered symbol can be covered by a new arrived encoding symbol in high probability. The ideal degree distribution for LRF codes is given by formula (5).

$$\rho'_d = \left(\frac{wn}{w^2-wn-n}\right)\left(\frac{1}{d(d-1)}\right) \quad d = \frac{w}{w-n}, \frac{w}{w-n}+1,\ldots,w. \quad (5)$$

Proof: The ideal soliton distribution in LT codes is $\rho_1,\ldots,\rho_w$, where

$$\begin{cases} \rho'_1 = \frac{1}{w} & d = 1 \\ \rho'_d = \frac{1}{d(d-1)} & d = 2,\ldots,w \end{cases} \quad (6)$$

Observe that in LRF codes, there are *n* output symbols with degree one already available in decoder. It do not need to generate encoding symbols with degree one. It has been concluded that the minimum degree is $\frac{w}{m}$. So the ideal degree distribution in dynamic window fountain codes is $\rho_{\frac{w}{m}},\ldots,\rho_w$, where

$$\rho_d = \alpha\left(\frac{1}{d(d-1)}\right) \quad d = \frac{w}{m}, \frac{w}{m}+1,\ldots,w. \quad (7)$$

It is obvious that $\sum_{d=\frac{w}{m}}^{w} \rho_d = 1$ is required by a probability distribution. So we get $\alpha = \frac{wn}{w^2-wn-n}$.

*C. Loss Rate Based Raptor Codes*

Raptor codes are the concatenation of high performance binary block code, such as LDPC and a weakened LT code. Hence, a loss rate based Raptor codes (LR-Raptor) may be obtained replacing the LT codes with the LRF codes. However, there are several specific features of Raptor codes which make this generalization not trivial. In Raptor codes, the degree distribution does not depend on the source symbol block length. In this subsection, a degree distribution with a max degree $d_{max}$ constraint is proposed. The degree distribution for LR-Raptor codes is given by formula (8) and (9).

$$\rho''_d = \beta\left(\frac{1}{d(d-1)}\right) \quad d = \frac{w}{w-n}, \frac{w}{w-n}+1,\ldots,d_{max}. \quad (8)$$

$$\beta = 1 / \sum_{d=\frac{w}{w-n}}^{d_{max}} \frac{1}{d(d-1)}. \quad (9)$$

*D. Performance Analysis*

In this subsection, a theoretical analysis of the properties of ideal degree distribution is provided. The following proposition shows that the decoding graph of a reliable decoding algorithm has at least an order of *m* encoding symbols.

*Proposition 1*: If a LRF codes with *w* input symbols and *n* output symbols with degree one possesses a reliable decoding algorithm, then there is a constant *c* such that the graph associated to the decoder has at least *m* output symbols with average degree $d_{avg}$. The *m* is given by formula (10).

$$m = \frac{cw}{d_{avg}} \ln\frac{w}{d_{avg}}. \quad (10)$$

*Proof*: In the decoding graph, if an input symbol has at least one output symbol as its neighbor, it is covered. Otherwise, it is uncovered. The failure probability of the decoder is lower-bounder by the probability that there is an uncovered symbol. We will establish a relationship between the numbers of output symbols and the average degree of an output symbol. The average degree of an output symbol in LRF codes is given by formula (11).

$$d_{avg} = \sum_{d=\frac{w}{w-n}}^{d_{max}} d\rho_d \quad d = \frac{w}{w-n}, \frac{w}{w-n}+1,\ldots,d_{max}. \quad (11)$$

Let *G* denote the decoding graph of the algorithm. It is a random bipartite graph between *w* input symbol and *n+m'* output symbols (*n* output symbols with degree one and *m'* output symbols with average degree $d_{avg}$ ). Let $V = \{v_1, v_2,..., v_m\}$ be the input symbols in $G$ which is not covered by $n$ output symbols. If an output symbol is of degree $d$, then the probability that $\frac{md}{w}$ symbol in $V$ is covered by this output symbol.

If an LT codes with $m$ input symbols possesses a reliable decoding algorithm, then there is a constant $c$ such that the graph associated to the decoder has at least $cm\ln m$ edges. Since the $\frac{md}{w}$ symbols in $V$ is covered by an encoding output symbol, a symbol in $V' = \{v_1, v_2,..., v_{m''}\}$ must be covered by an encoding output symbol where $m'' = \frac{kw}{md}$. As shown in Fig. 6, if 0.5 input symbols in $V$ ($k=2$) is covered by an encoding symbol, then an input symbol in $V'$ ($k=4$) is covered by an encoding symbol where other two input symbols in $V'$ is virtual. If two input symbols in $V$ ($k=4$) is covered by an encoding symbol, then an input symbol in $V'$ ($k=2$) is covered by an encoding symbol where two input symbols in $V$ is merged and became a input symbol in $V'$.

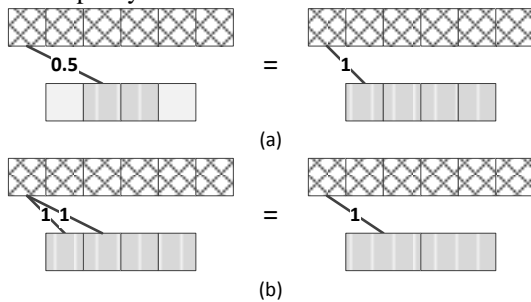

Fig. 6. Input symbols covered by encoding output symbols

## V. PERFORMANCE EVALUATION

In this section, several experiments are conducted to evaluate the performance of the proposed LRF codes. The performance analysis of LRF codes is performed through simulations varying the window length and lost rate. Thus, the simulation results have been proposed as a function of these parameters, while all the others were kept constant. Two previous digital fountain codes LT and Raptor codes are applied for comparisons. Simulations have been performed on a PC server, and the information of hardware and OS is shown in TABLE Ⅰ.

TABLE I. INFORMATION OF EXPERIMENT

| Host | CPU | Memory | OS |
|---|---|---|---|
| Source | 2*2.3GHz 8*core | 64GB | Linux 2.6.18 |

### A. Performance Analysis Varying Window Length

The performance results are expressed in terms of the encoding symbol ratio, i.e. the ratio of encoding symbols to lost symbols, and degree ratio, i.e. the ratio of degree to lost symbols. Both metrics are expressed as a function of window length in the range $[1\cdot10^3, 1\cdot10^5]$. The average encoding and decoding time per lost symbol are also recorded during the experiments.

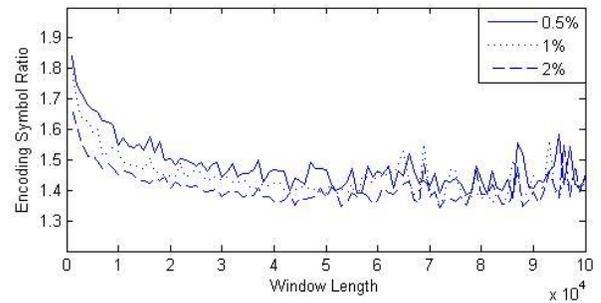

Fig. 7. Encoding Ratio as a function of the Window Length

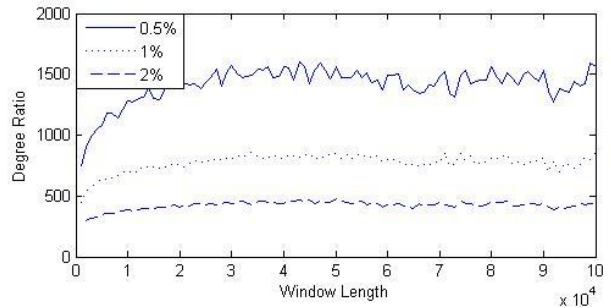

Fig. 8. Degree Ratio as a function of the Window Length

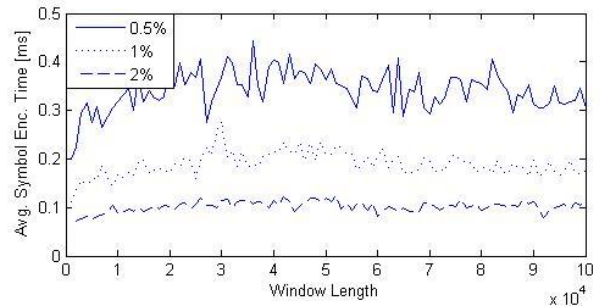

Fig. 9. Encoding Time as a function of the Window Length

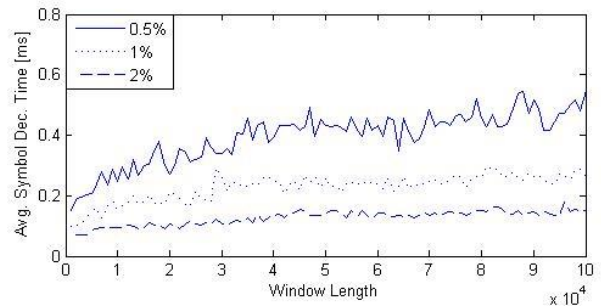

Fig. 10. Decoding Time as a function of the Window Length

Encoding symbol ratio versus window length with lost rate 0.5%, 1% and 2% refers to LRF codes is shown in Fig. 6. It can be noticed that bigger window length outperforms smaller window length in terms of encoding symbol ratio while window length less than $2\cdot10^4$. Degree ratio versus window length with lost rate 0.5%, 1% and 2% refers to LRF codes is shown in Fig. 7. It can be noticed that smaller window length outperforms bigger window length in terms of degree ratio while window length less than $2 \cdot 10^4$, and it also can be noticed that degree ratio in less lost rate is bigger. There is a conflict between encoding ratio and degree ratio. Average encoding and decoding time per lost symbol as functions of window length and loss rate are shown in Fig. 8 and Fig. 9. It can be noticed that LRF codes exhibits a linear encoding and decoding complexity, and the decoding time is increased along with the window length.

*B. Comparing With LT Codes*

The choice of the LT Robust Soliton distribution parameters, δ and *c*, are set in order to achieve quite good average performance. TABLE Ⅱ summaries all the parameters considered in the simulations. The performance results are expressed in terms of the encoding symbol ratio per input symbol, i.e. the ratio of encoding symbols to all input symbols, and degree ratio per input symbol, i.e. the ratio of degree to all input symbols. Both metrics are expressed as a function of loss rate in the range [0.1%, 5%]. The total encoding and decoding time are also recorded during the experiments.

TABLE II. INFORMATION OF EXPERIMENT

| LT Distribution parameters | δ | 0.5 |
|---|---|---|
| | *c* | 0.1 |
| Total Symbols | *t* | $1 \cdot 10^6$ Symbols |
| Symbol length | *l* | 1 bit |
| Window length | *w* | 10267 |

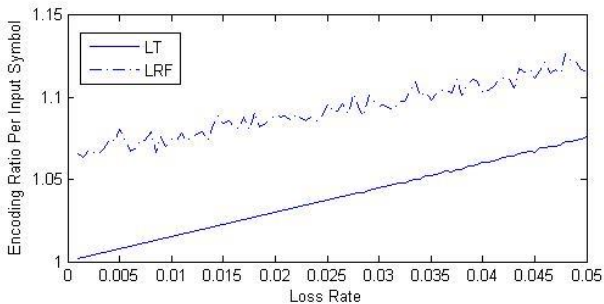

Fig. 11. Encoding Symbols as a function of the Loss Rate

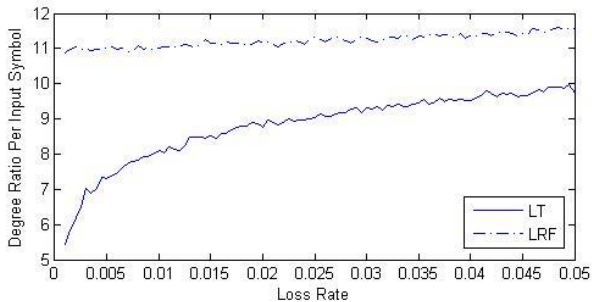

Fig. 12. Degree as a function of the Loss Rate

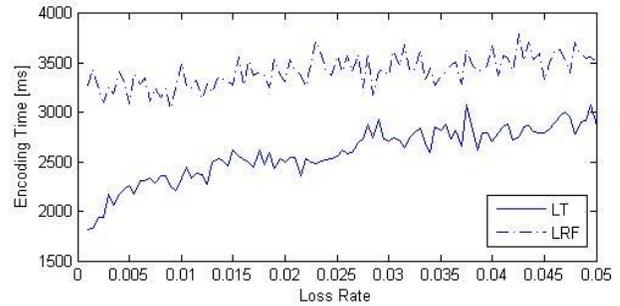

Fig. 13. Encoding Time as a function of the Loss Rate

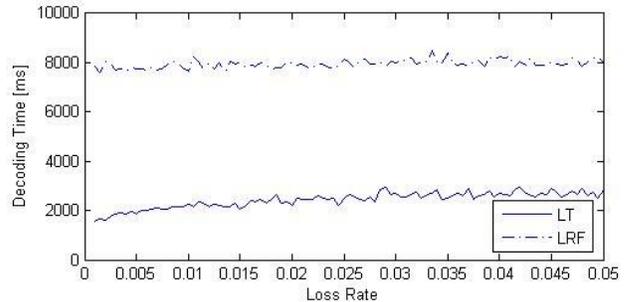

Fig. 14. Decoding Time as a function of the Loss Rate

Fig. 11 represents the encoding ratio as a function of the loss rate of LT and LRF codes. It can be notice that the performance of LRF codes is better than LT codes, and encoding ratio in LRF codes increase slight quicker than in LT codes. This is because LRF is approach to LT codes with the growth of loss rate. In LT codes, there are enough encoding symbols with degree one. When lots of input symbols are lost, LRF codes is similar with LT codes. Fig. 12 represents the degree ratio as a function of the loss rate of LT and LRF codes. It can be notice that degree ratio is grown slower than encoding ratio with the growth of loss rate. This is because the optimal degree value based on loss rate reduces the redundancy degree. Fig. 14, 15 represents the encoding and decoding time as a function of the loss rate of LT and LRF codes. It can be notice that both encoding and decoding time is lower in LRF codes. This is because there are less encoding symbols and less total degree in LRF codes.

*C. Comparing With Raptor Codes*

The choice of the Raptor parameters, *K*, *S* and *H*, are also set in order to achieve quite good average performance. TABLE Ⅲ summaries all the parameters considered in the simulations. The performance results are expressed as same as in comparing with LT codes.

TABLE III. INFORMATION OF EXPERIMENT

| | *k* | 10017 |
|---|---|---|
| Raptor codes parameters | *s* | 241 |
| | *h* | 11 |
| Total Symbols | *t* | $1 \cdot 10^6$ Symbols |
| Symbol length | *l* | 1 bit |

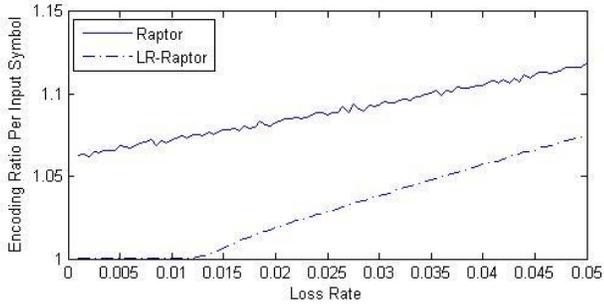

Fig. 15. Encoding as a function of the Loss Rate

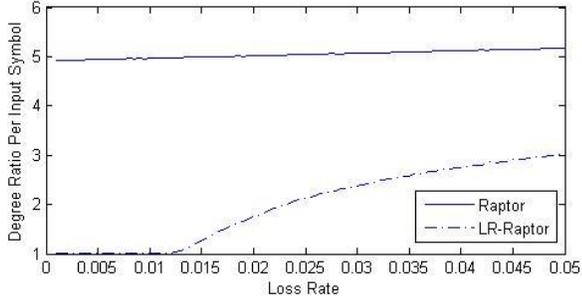

Fig. 16. Degree as a function of the Loss Rate

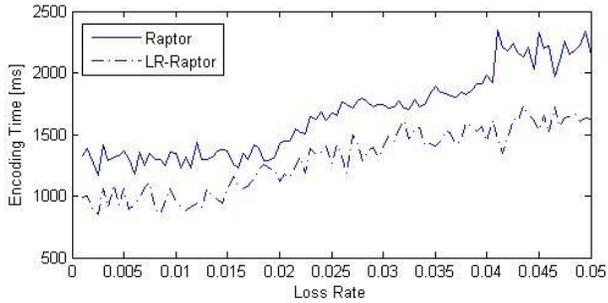

Fig. 17. Encoding Time as a function of the Loss Rate

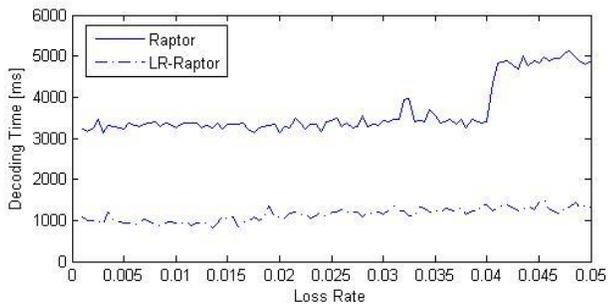

Fig. 18. Decoding Time as a function of the Loss Rate

Fig. 15 represents the encoding ratio as a function of the loss rate of Raptor and LR-Raptor codes. It can be notice that the encoding ratio in LR-Raptor codes is very small when loss rate less than 0.015, and encoding ratio in LR-Raptor codes is approach to Raptor codes with the growth of loss rate. This is because when little symbols are lost, most of input symbols have already been covered by output symbols with degree one, and only little encoding symbols are generated to cover lost symbols in LR-Raptor codes, While lots of encoding symbols are generated in Raptor codes. Fig. 16 represents the degree ratio as a function of the loss rate of Raptor and LR-Raptor codes. It can be notice that the degree ratio in LR-Raptor codes is very small when loss rate less than 0.015, and degree ratio in LR-Raptor codes is approach to Raptor codes with the growth of loss rate. However, degree ratio is grown slower than encoding ratio with the growth of loss rate. This is because, LR-Raptor make use of loss rate information and select optimal degree value. Fig. 17, 18 represents the encoding and decoding time as a function of the loss rate of Raptor and LR-Raptor codes. It can be notice that both encoding and decoding time is lower in LR-Raptor codes. This is because there are less encoding symbols and less degree in LR-Raptor codes.

### D. Comparing With LT and Raptor Code in Data Transfer

In this experiment, the performance of LRF codes by comparing with LT and Raptor codes for transferring large data is analyzed. All parameters are the same with experiment before. The size of data need to be transferred from source to destination is 4GB. The data is divided into 10 blocks, and each block has 10000 symbols of size 4096 bits. Symbols are sent from source to destination with a fixed rate 112MB/s. Netem is used in Linux to control the loss rate of the transmission channel. In order to full-utilize the CPU resource, Tools are implemented in multi-thread manner and data can be calculated and transferred concurrently. There are four threads taking care of data transfer and one thread takes care of program managing and scheduling. The information of hardware and OS is shown in TABLE Ⅳ. The bandwidth of network between source host and destination host is 1 Gb/s.

TABLE IV. INFORMATION OF EXPERIMENT

| Host | CPU | Memory | OS |
| --- | --- | --- | --- |
| Source | 2*2.3GHz 8*core | 64GB | Linux 2.6.18 |
| Destination | 2*2.3GHz 8*core | 64GB | Linux 2.6.18 |

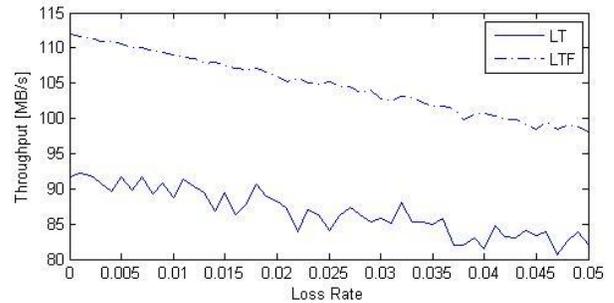

Fig. 19. Throughput as a function of the Loss Rate compare with LT

Fig. 19 represents the throughput as a function of the loss rate of LT and LRF codes. It can be notice that the throughput in LRF codes is bigger than in LT codes. This is because there are little redundancy encoding symbols in LRF codes, and LT codes take more CPU time than LRF codes which also reduce the throughput. Fig. 20 represents the throughput as a function of the loss rate of Raptor and LR-Raptor codes. It can be notice that the throughput in LR-Raptor codes is bigger than in Raptor codes. But the gap between Raptor and LR-Raptor codes is smaller than gap between LT and LRF codes. This is because Raptor codes is more efficient than LT codes.

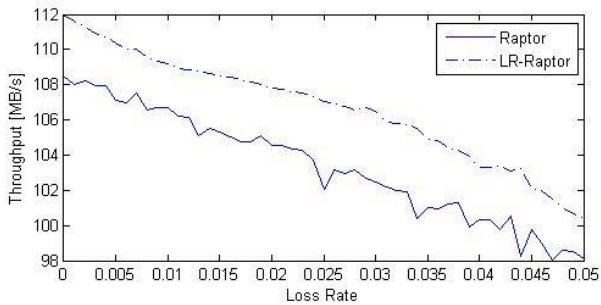

Fig. 20.  Throughput as a function of the Loss Rate compare with Raptor

## VI. CONCLUSION

In this paper, a novel fountain codes named LRF codes is presented. LRF codes provide an attractive alternative degree distribution for greatly reducing the running time and memory requirements for both encoding and decoding process comparing with LT codes and Raptor codes. Experimental evaluations show that LRF codes consistently outperforms LT codes and Raptor codes with respect to encoding ratio, degree ratio and encoding and decoding time. Because LRF codes make use of the information of loss rate and select optimal degree distribution, so it only need to transmit slight more encoding symbols instead of retransmitting lost symbols. So it is efficient for data transfer in high-speed long-distance network. Numerous challenges remain to be addressed. Our future work will mainly to reduce the redundancy rate of LRF codes and look at the applied technology with LRF codes for large data transfer. We intend to implementation tools for large data transfer with LRF codes and UDP.

ACKNOWLEDGMENT

This work was jointly supported by: (1) National Basic Research Program of China (No. 2013CB329102); (2) National Natural Science Foundation of China (No. 61271019, 61101119, 61121001, 61072057, 60902051); (3) PCSIRT (No. IRT1049).

REFERENCES

[1] I. Rhee, and L. Xu, "CUBIC: A New TCP-Friendly High-Speed TCP Variants," In Proceeding of PFLDnet, Lyon, France, February 2005.

[2] M. Bogino, P. Cataldi, M. Grangetto, E. Magli, and G. Olmo, "Sliding-Window Digital Fountain Codes for Streaming of Multimedia Applications," In Proceeding of SCAS, New Orleans, USA, May 2007.

[3] M. Luby, "LT Codes," In Proceeding of the ACM Symposium on Foundations of Computer Science, Vancouver, BC, CA, November 2002.

[4] A. Shokrollahi, "Raptor Codes," Information Theory, vol. 52, pp. 2551-2567, June 2006.

[5] P. Elias, "Coding for two noisy channels, Information Theory," Third London Symposium, Buttersworth's Scientific Publication, pp. 61-76, September 1955.

[6] J. Byers, M. Luby, M. Mitzenmacher, and A. Rege, "A Digital Fountain Approach to Reliable Distribution of Bulk Data," In Proceeding of SIGCOMM, Vancouver, BC, CA, September 1998.

[7] M. Luby, M. Mitzenmacher, A. Shokrollahi, and D. Spielman, "Efficient erasure correcting codes," IEEE Trans. Inform. Theory, vol. 47, pp. 569–584, 2001.

[8] 3GPP TS 26.346 V6.1.0, Technical Specification Group Services and System Aspects; Multimedia Broadcast/Multicast Service; Protocols and Codecs, June 2005.

[9] B. Sivasubramanian and H. Leib, "Fixed-rate Raptor codes over Rician fading channels," IEEE Trans. Vehicular Tech., Vol. 57, pp. 3905-3911, November 2008.

[10] G. Liva, E. Paolini, and M. Chiani, "Performance versus overhead for fountain codes over $F_q$," IEEE Commun. Lett., vol. 14, pp.178-180, February, 2010.

[11] K. Kasai, D. Declercq, and K. Sakaniwa, "Fountain Coding via Multiplicatively Repeated Non-Binary LDPC Codes," IEEE Trans. Commun., vol.60, pp.2077-2083, 2012.

[12] N. Rahnavard, B.N. Vellambi, and F. Fekri, "Rateless Codes With Unequal Error Protection Property," IEEE Trans. Inform. Theory, vol. 53, pp.1521-1532, 2007.

[13] K.C. Yang, and J.S. Wang, "Unequal Error Protection for Streaming Media Based on Rateless Codes," IEE Trans. Computers, vol.61, pp.666-675, 2012.

[14] P. Cataldi, M. Grangetto, T. Tillo, E. Magli, G. Olmo, "Sliding-window raptor codes for efficient scalable wireless video broadcasting with unequal loss protection," IEEE Transactions on Image Processing, vol. 19, pp. 1491-1503, June, 2010.

[15] C. Studholme and I. Blake, "Windowed erasure codes," in Proc. IEEE Int. Symp. Inform. Th., pp. 509-513, Seattle, WA, June, 2006.

[16] D. Sejdinovic, D. Vukobratovic, A. Doufexi, V. Senk, and R. Piechocki, "Expanding window fountain codes for unequal error protection," IEEE Trans. Commun., vol. 57, pp.2510-2516, 2009.

[17] M. Luby, "Raptor Forward Error Correction Scheme for Object Delivery," IETF, RFC 5053, Experimental Standard, October 2007.